\documentclass[aip,pop,reprint,groupaddress,noshowpacs,noshowkeys]{revtex4-1}

\usepackage{amsmath,amssymb,mathrsfs,multirow}
\usepackage{makecell}

\usepackage{graphicx}% Include figure files
\usepackage{dcolumn}% Align table columns on decimal point
\usepackage{bm}% bold math
\usepackage{xcolor}
\usepackage[pdftex,
  hidelinks,
  colorlinks=true,
  linkcolor=black,
  anchorcolor=black,
  citecolor=black,
  urlcolor=black,
  pdfauthor= {William Lewis},
  pdfsubject = {MagLIF Fuel Conditions},
  pdfdisplaydoctitle,
  bookmarks=true,
  bookmarksopen=true]{hyperref}

\begin{document}

\title{A framework for experimental-data-driven assessment of Magnetized Liner Inertial Fusion stagnation image metrics}
\author{William E. Lewis}
\email{willewi@sandia.gov}
\author{Eric C. Harding}
\author{David A. Yager-Elorriaga}
\author{Jeffrey R. Fein}
\author{Patrick F. Knapp}
\altaffiliation[Current address ]{Los Alamos National Laboratory}%Lines break automatically or can be forced with \\
\author{Kristian Beckwith}
\author{David J. Ampleford}

\affiliation{Sandia National Laboratories, Albuquerque, New Mexico 87185 USA}

\date{\today}

\begin{abstract}
A variety of spherical crystal x-ray imager (SCXI) diagnostics have been developed and fielded on Magnetized Liner Inertial Fusion (MagLIF) experiments at the Sandia National Laboratories Z-facility. These different imaging modalities provide detailed insight into different physical phenomena such as mix of liner material into the hot fuel, or cold liner emission, or reduced impact of liner opacity. However, a number of practical considerations ranging from the lack of a consistent spatial fiducial for registration to different point-spread-functions and tuning crystals or using filters to highlight specific spectral regions make it difficult to develop broadly applicable metrics to compare experiments across our stagnation image database without making significant unverified assumptions. We leverage experimental data for a model-free assessment of sensitivities to instrumentation-based features for any specified image metric. In particular, we utilize a database of historical and recent MagLIF data including $N_{\text{scans}} = 139$ image plate scans gathered across $N_{\text{exp}} = 67$ different experiments to assess the impact of a variety of features in the experimental observations arising from uncertainties in registration as well as discrepancies in signal-to-noise ratio and instrument resolution. We choose a wavelet-based image metric known as the Mallat Scattering Transform for the study and highlight how alternate metric choices could also be studied. In particular, we demonstrate a capability to understand and mitigate the impact of signal-to-noise, image registration, and resolution difference between images. This is achieved by utilizing multiple scans of the same image plate, sampling random translations and rotations, and applying instrument specific point-spread-functions found by ray tracing to high-resolution datasets, augmenting our data in an effectively model-free fashion. We anticipate that our work will contribute to enabling strand morphology to inform inferences of physical conditions.\end{abstract}



\maketitle

\section{\label{sec:intro}Introduction}

Magnetized Liner Inertial Fusion (MagLIF)~\cite{Slutz_PoP_2010} is a magneto-inertial fusion concept being studied at the Sandia National Laboratories Z-facility. Fusion relevant conditions are achieved by compressing a premagnetized and preheated deuterium fuel contained in a cylindrical beryllium tube or liner with the magnetic pressure generated by an $\mathcal{O}$($20$ MA) current. Premagnetization with an axially oriented magnetic field of $\mathcal{O}$($10$ T) is provided by external field coils~\cite{Rovang_RSI_2014} and the fuel is preheated by a kilo-Joule-class laser.~\cite{Rambo_AO_2005,Rambo_SPIE_2016} As the liner is accelerated inward due to the high magnetic pressure generated by the drive current, the fuel is quasi-adiabatically heated through $PdV$ work with magnetic flux compression providing for reduction in thermal conduction losses and trapping of charged fusion products. The implosion decelerates when the thermal pressure in the fuel becomes great enough to balance the magnetic drive pressure, eventually resulting in stagnation with ion temperatures $T_{\text{ion}} \approx 3$~keV, fuel radial areal densities $(\rho R)_{\text{fuel}} \approx 2$~mg/cm$^2$, and several kilo-Tesla magnetic fields.~\cite{Gomez_PRL_2014}

The self-emission x-rays from the stagnated fuel contains critical information for diagnosing plasma conditions and morphology. One approach to accessing this information experimentally is through the application of  a variety of different spherical crystal x-ray imaging (SCXI) capabilities.~\cite{Harding_Prep} Single-, dual-, and orthogonal- imaging modalities have been fielded on that include the possibility for spectral filtering to focus on spectral regions which contain information that continues to shape our understanding of quantities such as mix, liner areal density, and morphology of the stagnated fuel. Unfortunately, extracting maximum value from these datasets has remained a challenge due to significant expertise and manual input required to process and analyze the images and uncertainty over the comparability of images collected with different imaging diagnostics that differ in instrument response. As a result, it has generally only been feasible to compare images coming from a small number of experiments specifically designed with the intent of showing gross changes in the overall structure of the plasma.~\cite{Gomez_PRL_2020, Glinsky_PoP_2020, Harding_Prep, Ampleford_Prep} 

In a companion article to this, we demonstrated the development of a deep-learning-based automated processing routine to enable the preparation of a large dataset consisting of nearly all usable MagLIF SCXI stagnation images.~\cite{Lewis_JPP_2022} In this manuscript, we leverage that dataset to demonstrate a model-free approach to understanding the impact of various instrument response features as well as sensitivity to image registration. In Sec.~\ref{sec:Metrics}, we provide brief motivation for using a fixed-weight convolutional network known as the Mallat Scattering Transform (MST)~\cite{Bruna_IEEE_2013} to quantify our images. However, we highlight that the approach of comparing images directly in the chosen metric space pursued in Sec.~\ref{sec:Results} allows us to assess sensitivities of the metric to realistic features of experimental data is valid independent of the choice of image metric. The observed sensitivities may be contingent upon a variety of factors, for example, metric choice, particular diagnostic configuration, as for images taken at nearly the same view angle but with different spectral filtering or for optical setups with different imaging resolution or point-spread-functions (PSFs), and data digitization properties such as scanner resolution, gain settings, or effects of multiple scans of the same image plate. In addition, the structure of the stagnation column should also play a role in determining the image metric. We find that the MST exhibits undesirable sensitivity to statistical ``texture" introduced by digitization of the image plate. As a result, multiple scans of the same image plate exhibiting different signal-to-noise ratios (SNRs) but otherwise identical structure can exhibit separation in the metric space of a scale similar to that between images from different experiments. We demonstrate a ``texture-subtraction" method to remove the impact of this effect. Next we demonstrate that the MST does exhibit some sensitivity to image registration. However, due to the fact that the MST varies slowly with translation and rotation, these transformations result in a roughly linear response that we show can be effectively projected out. We also demonstrate the extent to which thee image metric is sensitive to differences in resolution. In particular, we show how high-resolution SCXI data can be augmented using PSFs computed using ray tracing to understand the impact this has on image comparison. Finally, in Sec.~\ref{sec:morphology}, we demonstrate that the resulting metric appears to be sensitive to qualitatively apparent differences in strand morphology. 

\section{\label{sec:Metrics} Image Metrics}
Unfortunately, a variety of factors have played a role in preventing large-scale simultaneous analysis of SCXI data. Perhaps the most important of these factors include: the time consuming processing methods which may vary for each investigator, lack of data management best practices for reproducibility, and difficulty in defining and understanding image metrics that allow for direct \textit{quantitative} comparison of stagnation images that may have been fielded using quite different imaging modalities and/or spectral filtering as detailed in the previous section. As a result, studies of the spatial structure of the stagnated fuel-liner system have typically target between one and a few images from a particular shot or experimental campaign. While the resulting studies have successfully revealed important physical insights, restricting analysis to a small number of experiments inherently risks biasing analysis through unintended ``cherrypicking", or simply missing deeper insights that may be buried in our datasets. By improving automation of image processing and recording relevant metadata in Ref.~\onlinecite{Lewis_JPP_2022}, we have opened up the possibility to apply image metrics across our entire dataset, enabling large-scale studies. 

Here we focus on understanding how to utilize our experimental data to design an image metric that accurately accounts for realistic features arising \textit{e.g.} from instrument response. In order to undertake a sensitivity study, it is necessary to choose a particular method for quantifying images. A number of reasonable approaches have been applied, or are under current investigation. For example, in Ref.~\onlinecite{Harding_Prep}, for experiments designed to look at the structure of mix from the liner wall into the fuel, registering and subtracting images from each channel of a dual-crystal imager configuration revealed structural information about wall mix. In Ref.~\onlinecite{Ampleford_Prep}, a Fourier transform was applied to the axial strand intensity profile after integrating along horizontal line outs as well as to the strand radial amplitude in order to characterize structure. In Ref.~\onlinecite{Glinsky_PoP_2020}, a wavelet-based transform was paired with regression onto a set of geometric parameters used to generate two-dimensional synthetic images to extract interpretable morphological features. Each of these approaches has several benefits as well as drawbacks. The first approach reveals rich details in the spatial structure of the physics under investigation. However, it is only applicable to individual experiments with pairs of DCI images as it offers no metric for experiment-to-experiment comparison. In addition, uncertainty caused by sensitivity to registration of image pairs may be difficult to address. The second approach provides a highly intuitive set of hand-picked features by decomposing the image into the spectral content of two traces of helical excursion amplitude and radially integrated intensity versus axial position. However, the resulting Fourier-spectrum is not very sparse, so peak identification and spectral comparison are rather difficult and subject to potential human bias in all but the simplest of inferences. The third approach utilizes a wavelet-based metric that can be computed for any given image, is fairly interpretable. The wavelet-based approach offers some desirable mathematical guarantees. However, it is unclear that the proposed background subtraction and extraction of morphological parameters based on the two-dimensional geometric model are directly comparable, especially between different experiments with different image modalities and spectral filtering. As a result, the application of this method is fairly restricted. Furthermore, the simple model likely doesn't characterize all of the features in our experimental datasets, which may result in uncontrolled bias in the parameter inference. Our work here is complimentary to all three of these approaches by considering the quantification of a much large dataset than has previously been undertaken. It also provides a means by which one can assess the sensitivity of each of these approaches to differences arising from registration and instrument response. 

\subsection{\label{sec:MST} Wavelet-based image metric}
We choose to quantify images using a wavelet-based image metric known as the Mallat Scattering Transform (MST). In this section, we provide a brief self-contained overview and intuition for this metric, but the reader is referred to some of the many works that detail and apply this and related metrics.~\cite{Bruna_IEEE_2013, Glinsky_PoP_2020,  Allys_PRD_2020} The MST converts the spatial information to spatially localized inverse scale, \textit{i.e.} spatial-frequency, and cross-scale correlation (\textit{i.e.} coherence) information through a cascaded application of band-pass filtering, non-linear rectification, and a pooling or averaging operation. The band-pass filters are indexed by two integers $j,\ell$, which characterize the transformation of a ``mother wavelet" $\psi$ to give the $j-\ell$ band-pass filter
\begin{eqnarray}
	\psi^\ell_j(\mathbf{x}) = 2^{-j} \psi(2^{-j} r_{\ell} \mathbf{x}).
\end{eqnarray}
In the above equation, $j \in \{0,1,...,J-1\}$ sets the spatial frequency for the particular band-pass filter, with larger $j$ corresponding to lower frequency, while $\ell$ sets it's orientation, with $r_{\ell}$ specifying a coordinate rotation operation to set orientation of the filter to be $2\pi \ell/L$, where $\ell \in \{0,1,...,L-1\}$ with $L$ being the total number of orientations that the filter is allowed to take. We use the open source library \textit{kymatio}~\cite{Kymatio_Ref} in python for computing the MST using the default settings, for which $\psi(\mathbf{x})$ is taken to be a complex valued Morlet wavelet 
\begin{eqnarray}
	\psi(\mathbf{x}) = A (e^{i \mathbf{k}_0\cdot\mathbf{x}}-B)e^{\frac{-\mathbf{x}^2}{2 \sigma_0^2}},
\end{eqnarray}
where $A$ fixes $\int \psi^2(\mathbf{x}) d\mathbf{x} = 1$, $B$ fixes $\int \psi(\mathbf{x}) d\mathbf{x} = 0$, t he spatial frequency is centered near $k_0 = 3 \pi/4$, and $\sigma = 0.8 \approx \pi/4$ corresponding to a bandwidth $\delta_0 \approx \pi/2$ in dimesionless ``pixel-units". Note then that the center frequency and bandwidth of $\psi_j^\ell$ are $k_j \approx k_0/2^j$ and $\delta_j \approx \delta_0/2^j$. 
The first order of the MST is given by
\begin{eqnarray}
	S^1_{j,\ell}(\mathbf{x}) = |\mathcal{I}(\mathbf{x}) \star \psi^\ell_j(\mathbf{x})|\star \phi_J(\mathbf{x})\label{eq:FOMST} ,
\end{eqnarray}
where $\mathcal{I}$ is the image being transformed, and $\phi_J$ is a Gaussian spatial window with width proportional to $2^{J-1}$. Notice that the width of this envelope function is scale-matched to the largest scale resolved by the family of mother wavelets. 

Intuitively, the modulus term in Eq.~\ref{eq:FOMST} can be interpreted as containing a contribution from the total energy contained in the band, along with beat-frequency interference between the different spatial frequencies contained within the band that share any local coherence. The largest such beat-frequency possible is the bandwidth of the bandpass filter under consideration, which is proportional $1/2^j$. The subsequent convolution with $\phi_J$ will thus tend to average out the interference term in the modulus, causing a loss of information about the coherence between spatial frequencies. A second layer of filtering may then be applied before convolution with $\phi_J$ to help recover this information
\begin{eqnarray}
	S^2_{j,\ell,j',\ell'}(\mathbf{x}) =\big||\mathcal{I}(\mathbf{x}) \star \psi^\ell_j(\mathbf{x})|\star \psi^{\ell'}_{j'}(\mathbf{x})\big|\star \phi_J(\mathbf{x}).
\end{eqnarray}
Since the modulus in the definition of $S^1$ produces low frequency structures that $S^2$ is meant to recover, we may restrict to the case $j' > j$. The cascading process may be continued, but the operation is contracting at each level due to propagating energy to ever lower frequency, causing the coefficients to shrink in magnitude.~\cite{Bruna_IEEE_2013} Thus we will only consider up to this ``second-order" scattering transform. In addition, we note that the averaging operator $\phi_J$ will tend to introduce strong correlations in the final output at each image location $\mathbf{x}$. Put another way, this contributes to reducing sensitivity to translation, but at the cost of producing redundant information. Particularly, for translation by a vector $\mathbf{c}$ with $|\mathbf{c}| < < 2^J$, the scattering coefficients are nearly unchanged. Rotations by $\Theta < <  2 \pi/L$ will not strongly impact the result. We are thus free to subsample the final convolution with $\phi_J$ at points $\mathbf{u}$ by setting an appropriate convolution stride based on the scale set by $2^J$ and our image dimensions as we will describe in Sec.~\ref{sec:tsne}. Furthermore, we note that small changes in image resolution will tend not to shift information out of a given frequency band, so that the metric is stable to slight changes in \textit{e.g.} resolution and magnification. More generally, the MST is said to be Lipschitz continuous to deformations, which means the metric will change proportionally to any deformations such as small contrast changes or small changes in image registration, making this a good choice for image-to-image comparison. One may anticipate that the slight differences in resolution between different SCXI modalities or the lack of fielding a spatial fiducial will not spoil our ability to use this metric to compare images. However, it is critical that the impact of these transformations as implemented by limitations of realistic experimental instrumentation be assessed in order to ensure accurate understanding when making systematic quantitative comparison of different images. We will demonstrate sensitivity to all of these features using only experimental datasets in Sec.~\ref{sec:Results}. 

By construction, the MST utilizes bandpass filters and a pooling function whose frequency and bandwidth scale as powers of $2$. For this reason, it is convenient to subsample images onto a grid with dimensions $2^K\times2^K$. We choose $K=9$, \textit{i.e.} all of our images are subsampled to $512 \times 512$ pixels before applying the MST. Unless otherwise noted, we let $J=7$ and $L=8$ resulting in the scale of $\phi_J \sim2^J$ being ~$1/4 \times 1/4$ of our image scale. In object-space units, the construction of the MST sets a non-square pixel size of about $5~\mu$m wide by $23~\mu$m tall with $\sim4.0^{+2}_{-1}$mm ($0.8^{+0.4}_{-0.2}$mm) as the longest and $62.5^{+31.3}_{-15.6}~\mu$m ($13.0^{+6.5}_{-3.3}~\mu$m) as the shortest wavelengths resolved by the bandpass filters in the vertical (horizontal) direction to be used in the MST. Note that if we chose a larger value of $K$ this would allow better resolution of smaller scale structures. However, we must simultaneously increase $J$ to keep the largest resolved scale from decreasing. With this perspective it is easy to see how the MST may be thought of as a fixed-weight convolutional network having an output at each layer, with each order of the MST corresponding to a network layer, the modulus operation acting as a nonlinear rectification, and the convolution with the father-wavelet and subsampling corresponding to pooling operations.~\cite{Bruna_IEEE_2013} In the next section we describe how the MST coefficients are visualized along with a brief discussion of other data-exploration methods we will use to assess sensitivity of the MST to features in our experimental images.

\subsection{\label{sec:tsne}Data visualization and dimensionality reduction}

\begin{figure}[ht!]
\includegraphics[width=0.5\textwidth]{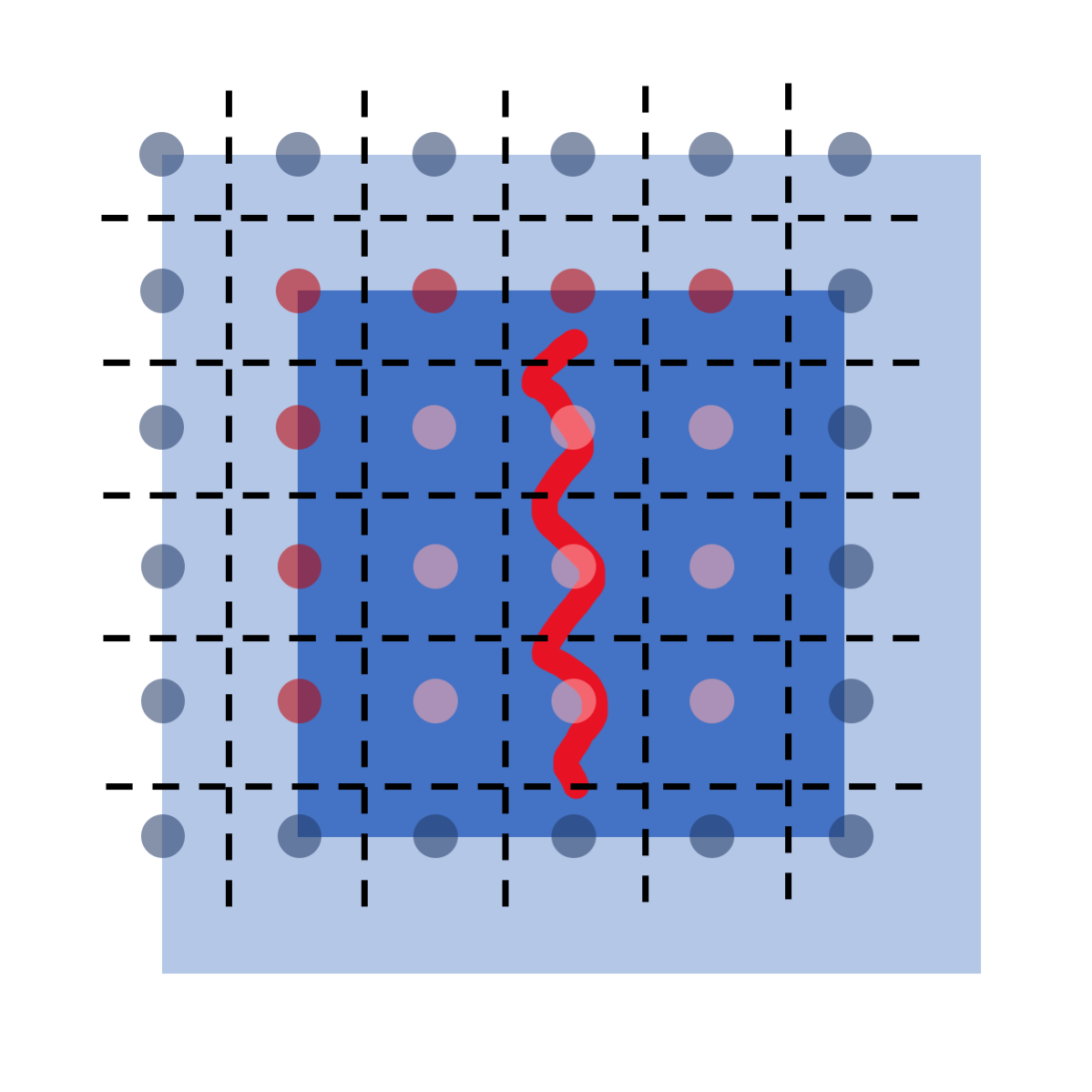}
\caption{To compute the MST coefficients, an image (dark blue) is reflection padded (light blue). Bandpass filtering, final convolution with $\phi_J$, and downsampling of coefficients to locations $\mathbf{u}$ are achieved in the Fourier domain. Coefficients sampled at the blue dots are discarded by the algorithm, which outputs coefficients at the red and pink locations. Because our signal exhibits significant spatial localization, only coefficients at the three pink dots in the central column of the original image are kept, while coefficients output the additional ``off-strand" pink dots are used to formulate one possible texture subtraction procedure in Sec.~\ref{sec:noise_sens}.}\
\label{fig:padding}
\end{figure}

We mentioned in Sec.~\ref{sec:MST} that $J$ sets the stride that should be chosen to determine the locations at which the MST coefficients will be output. In particular, to minimize redundant information the stride is selected so that each location $\mathbf{u}$ is separated by $2^J$ pixels resulting in outputting coefficients at $(N_\text{pix}/2^J)^2$ spatial locations to avoid redundant information. While we won't delve into the algorithmic details, \textit{kymatio} computes convolutions and downsampling corresponding to the stride of $\phi_J$ in the Fourier domain with the image being reflection padded to the necessary size for the computation. The exact locations at which coefficients are output are impacted by these choices, and is indicated by the pink and red dots visually in Fig.~\ref{fig:padding}. Utilizing the registration procedure of Ref.~\onlinecite{Lewis_JPP_2022}, the strands present in our images will primarily occupy the 3 vertical boxes in the center column primarily contain the most relevant information. However, we do note that the boxes are not to be interpreted as hard boundaries for spatial localization of information. Recall that the largest observable scale is $\sim 6$~mm in the vertical direction, while the box size is half that distance. As a result, information from nearby boxes will share some resemblance. In order to avoid poorly controlled boundary effects, we will keep only coefficients on the $3 \times 3$ grid shown by the pink dots in Fig.~\ref{fig:padding}. Note that we will also use the columns to the left and right, henceforth ``off-strand", later in the manuscript. 

\begin{figure}
\includegraphics[width=0.5\textwidth]{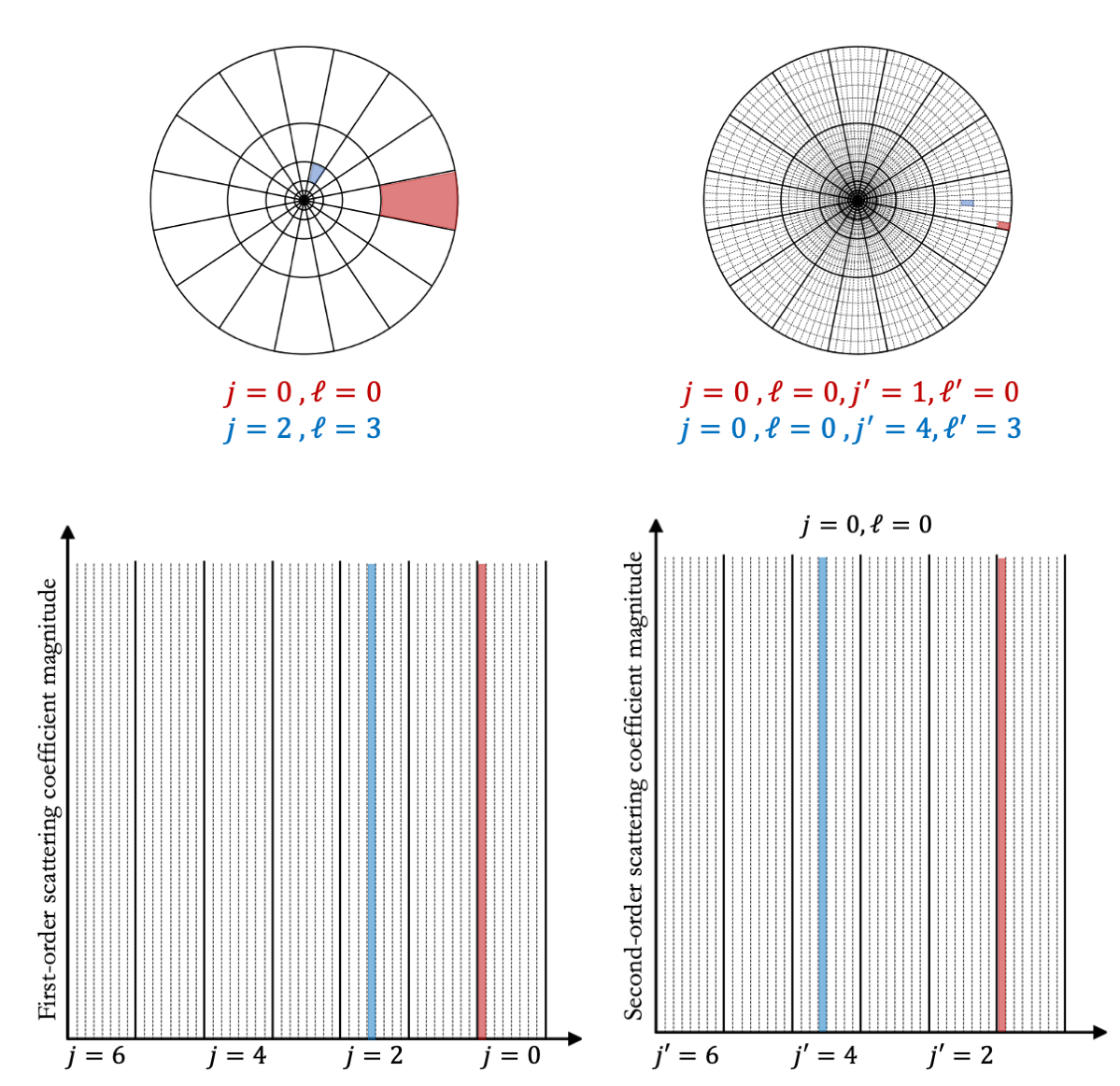}
\caption{Upper row: Scattering disk display for first- and second-order MST coefficients with $J=7$, $L=8$ demonstrating correspondence of bins in the display with MST indicies $j$, $\ell$, $j'$, and $\ell'$. Note that the colors here are meant to highlight a particular bin rather than represent a coefficient value in this case. When plotting scattering coefficients, color will indicate magnitude. Lower row: Structure for line-plots of scattering coefficients. Note that within each $j$ ($j'$) bin $\ell$ ($\ell'$) increases from $0$ to $7$. Also note that for the second-order coefficients, a separate line-plot would be required for each $j,\ell$ pair for $j <6$. For this reason, we will generally only show a small subset fo second-order coefficients in the form of line-plots.}
\label{fig:scatdisk}
\end{figure}

We emphasize that, at each of the pink spatial locations in Fig~\ref{fig:padding}, a set of $N_1 = JL$ first-  and $N_2  = J(J-1)L^2/2$ second-order coefficients are output, quantifying the energy in each spectral band and information about interferences within a given band respectively. This informs the two different ways in which MST coefficients will be visualized in this manuscript. The first of these is using the so-called a scattering-disk display shown in the upper row of Fig.~\ref{fig:scatdisk}.~\cite{Bruna_IEEE_2013} In this representation, the first- and second-order coefficients are plotted on two separate disks. The first order may be directly interpreted as displaying the energy content within a given frequency band $j$ with frequency increasing versus disk radius and bin size proportional to the bandwidth, while the filter angle $\ell$ is given by the angular coordinate. The second disk splits each of the original $j$ and $\ell$ bins up into a total of $(J-1)-j$ bins radially for the $j'$ coordinate, and a total of $L$ bins in the angular direction for the $\ell'$ coordinate. Note that in this case, for ease of viewing, each of the bins are of the same radial and angular size within a given $j-\ell$ bin. The second row in Fig.~\ref{fig:scatdisk} shows how the coefficients can be plotted on a series of line-plots. For this case, the frequency is increasing with each $j$ ($j'$) bin, while $\ell$ ($\ell'$) increases within each $j$ ($j'$) bin. 

Before continuing, we pause to note that there are several reasonable choices for normalizing images and/or corresponding MST coefficients. These could, for example, be one of the following
\begin{eqnarray}
	\overline{\mathcal{I}}(\mathbf{x}) &=& \dfrac{\mathcal{I}(\mathbf{x})-\mathcal{I}_\text{min}}{\mathcal{I}_\text{max}-\mathcal{I}_\text{min}} \text{\phantom{...}(Min-max norm image)},\\
	\overline{\mathcal{I}}(\mathbf{x}) &=& \dfrac{\mathcal{I}(\mathbf{x})}{\int  \mathcal{I} d\mathbf{x}} \text{\phantom{...}(Integral norm image)},\\
	\overline{S} \equiv (&\overline{S}^1,& \overline{S}^2) = \dfrac{S}{\sum S^1} \text{\phantom{...}(Spectral norm)}, \label{eq:specnorm}
\end{eqnarray}
where the sum in Eq~\ref{eq:specnorm} is taken to run over the all first order coefficients and for the column of interest in the image. Through exploration of our dataset similar to what we conduct in Sec.~\ref{sec:Results}, we found that the first two choices were not appropriate for our experimental dataset. The min-max and integral normalization choices are more susceptible to outlier bright and dark pixels from \textit{e.g.} plate damage or clipping and bright background emission. Hence in the following, we utilize the spectral normalization, dividing by the sum of the first-order MST coefficients. Note that this does not equate to the total spectral energy, as we have not accounted for the bandwidth of each spectral bin. However, that would tend to weight the higher-frequency bins with their greater bandwidth more significantly, while most of the relevant strand-information is contained in the lower frequency bins.

\begin{figure}
\includegraphics[width=0.45\textwidth]{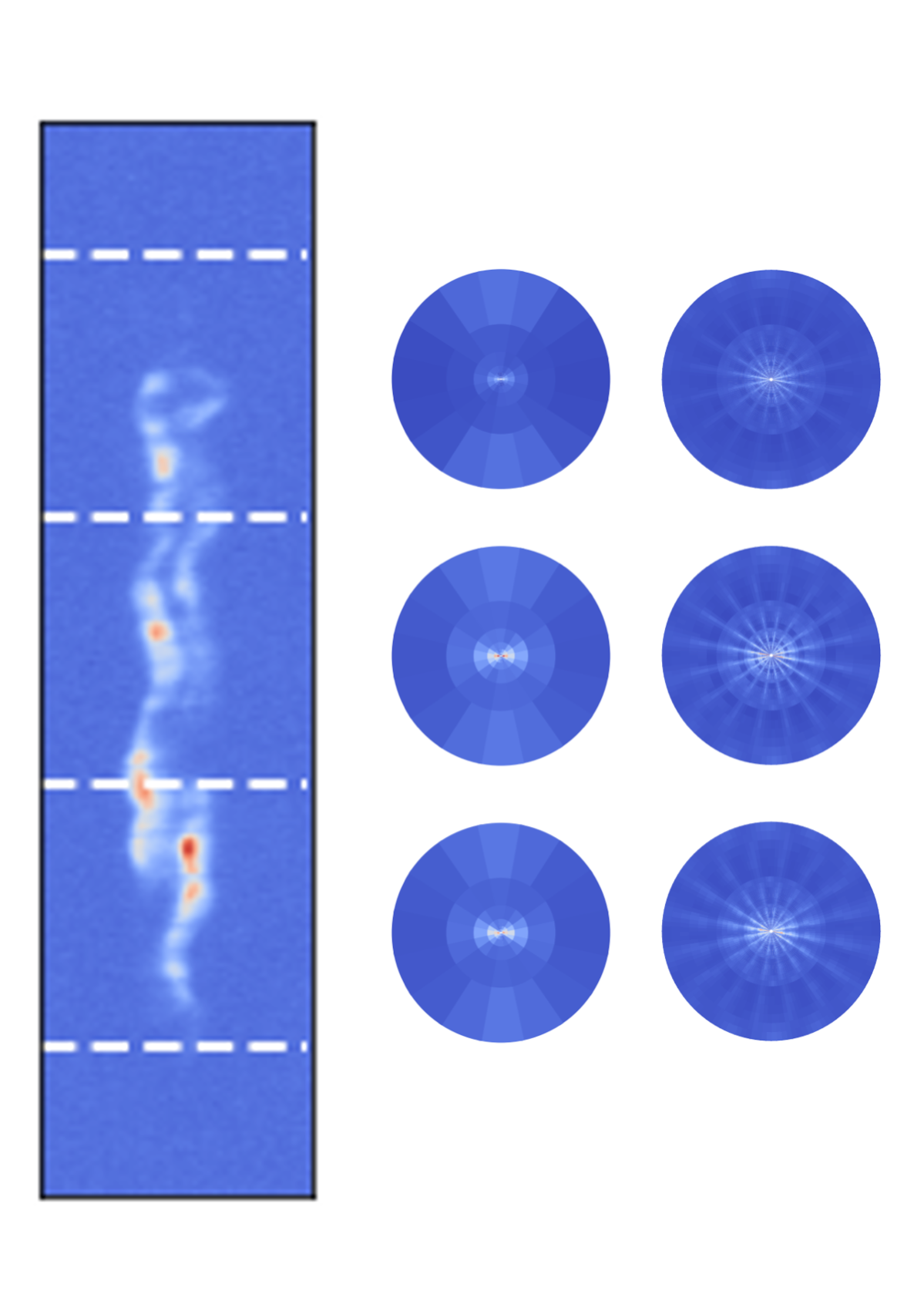}
\caption{Stagnation image and MST of z3564.}
\label{fig:MST_Experiment_Demo}
\end{figure}

Having defined the basic metric we will use to quantify images, we may now proceed to studying its application to experiment. Fig.~\ref{fig:MST_Experiment_Demo} demonstrates one such example, showing only the three scattering disks for each of the first- and second-order MST coefficients from the central column of the image containing. Of note, the MST provides a significant dimensional reduction of the image information. We note that there are a total of $N_\text{im} \sim 65$k pixels in the central column of the image. Supposing that only pixels corresponding to the strand contain relevant information, there's still about $\overline{N}_\text{im} \sim 16$k pixels. On the other hand, there are only $N_{\text{MST}} = 4.2$k MST coefficients corresponding to a significant dimensionality reduction factor of between $4-15$.

Despite the significant dimensionality reduction gained by using the MST, the metric space is still quite high-dimensional, so that visualizing and comparing data large amounts of data in the MST metric space will remain challenging. One simple way to visualize relationships is via the pairwise distance matrix 
\begin{eqnarray}
	D_{i,j} = \frac{||\mathbf{x}_i - \mathbf{x}_j||_2}{d_\text{norm}},
\end{eqnarray}
where $||\cdot||_2$ is the L$_2$ norm and $d_\text{norm}$ is a suitably chosen normalization constant. For this manuscript, we let $d_\text{norm}$ be given by the $50^\text{th}$ percentile of all the pairwise distances (including self distances $D_{i,i}$). While such a visualization will be useful when considering small subsets of the data or to pick out outliers, alternate visualization methods may be more insightful. In order to improve our ability to visualize the data, we utilize principal component analysis (PCA).~\cite{Bishop_2006} Briefly, we remind the reader that for a dataset $\mathbf{X} \in \mathbb{R}^{N_{\text{samps}} \times N_{\text{features}}}$, the principal components are the set of ordered orthogonal vectors describing the directions of greatest variance in the dataset. These are typically computed by singular value decomposition
\begin{eqnarray}
	\mathbf{X} = \mathbf{U}\mathbf{\Sigma}\mathbf{V}^T,
\end{eqnarray}
where $\mathbf{U}~\in~\mathbb{R}^{N_{\text{samps}}~\times~N_{\text{samps}}}$ is orthogonal, $\mathbf{\Sigma}~\in~\mathbb{R}^{N_{\text{samps}}~\times~N_{\text{features}}}$ is a diagonal matrix, and $\mathbf{V}~\in~\mathbb{R}^{N_{\text{features}}~\times~N_{\text{features}}}$ is orthogonal with the columns of $\mathbf{V}$ forming the principal vectors for the dataset $\mathbf{X}$. Considering the coefficients by projection onto the first few principal vectors is often used for dimensionality reduction. Here we will show plots of the first two principal components, but we not that in general, this will not be sufficient to draw detailed conclusions for example about pairwise distance between points. For example, we find that $N_{95\%} = 16$ principal components must be kept in order to explain $95\%$ of the dataset variance. This does however constitute a significant additional reduction in the dataset dimensionality, down from about $4.2$k coefficients down to about $16$ coefficients. Finally, we note that before computing principal components, the MST coefficients are standard normal scaled, so that each coefficient bin has zero mean and unit standard deviation across our entire dataset. This is done to ensure that the direction of maximum variance is not dominated by the bins with largest coefficient values. We also note that in finding the direction of greatest variance, PCA is inherently sensitive to outliers in the dataset. This will inform additional choices made Sec.~\ref{sec:noise_sens} regarding proposed ``texture subtraction'' methods.

\section{\label{sec:Results} Experiment-driven Metric Design}
\begin{figure}
\includegraphics[width=0.5\textwidth]{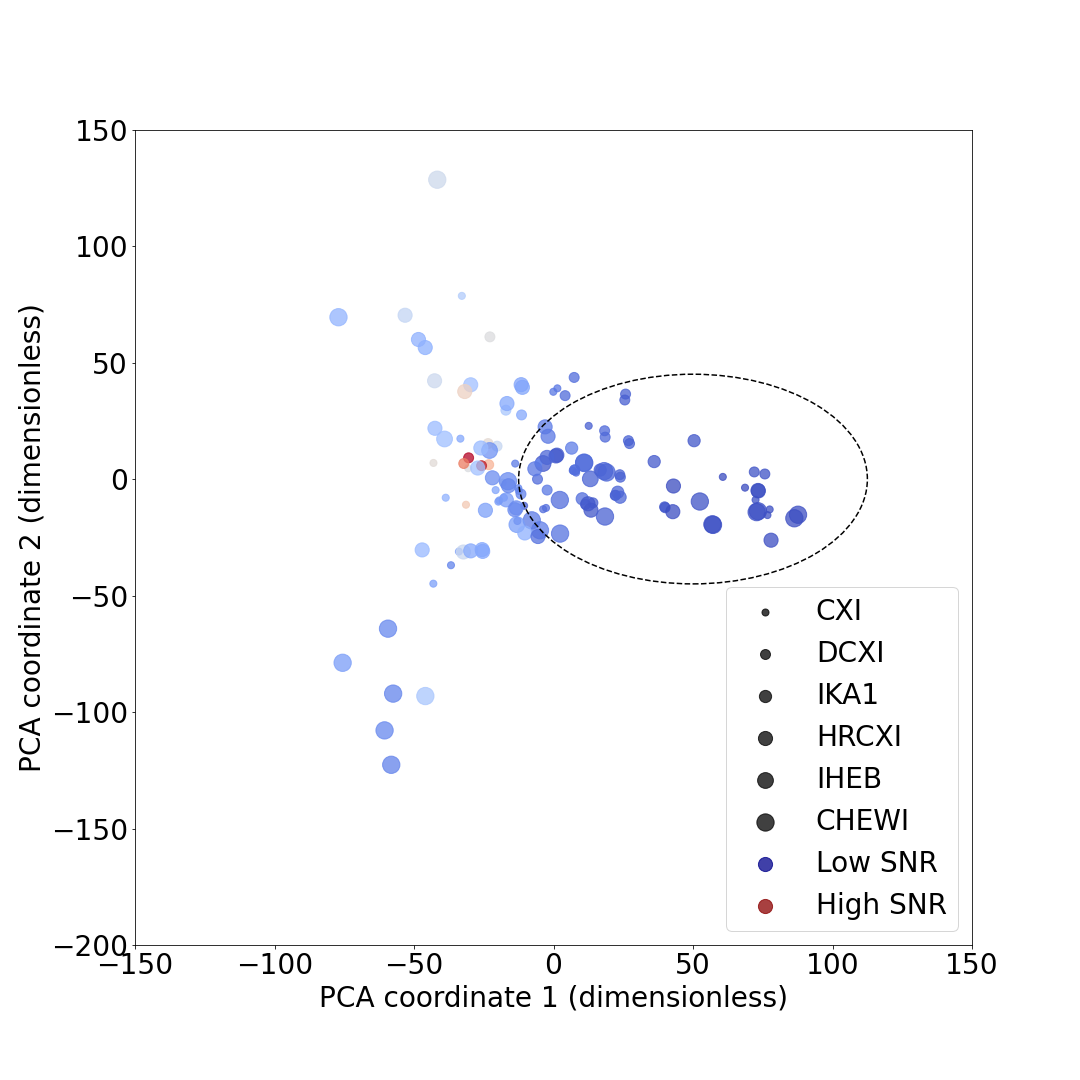}
\caption{The application of PCA to the MST computed for all of our experimental images appears to separate to some extent along signal-to-noise ratio, with the black dashed ellipse containing only lower SNR images. On the other hand, different modalities which will differ in spectral filtering and imager resolution appear to be well interspersed at least for the first two principal components indicating they aren't the dominant contribution to variance in the data.}
\label{fig:tSNE_All_Exp}
\end{figure}

In this section, we demonstrate how historical MagLIF data along with the above image metric and data visualization methods can be used to understand and reduce sensitivities to experimental features arising purely from instrument response and uncertainty in registration. It is expected that this will as a result improve sensitivities of the metric to physics-based features of our dataset that will also be explored at the end of this section. We begin with dataset exploration. We use the processing method presented in Ref.~\onlinecite{Lewis_JPP_2022} to register images and subtract off a slowly varying background (fit with a polynomial form). Fig.~\ref{fig:tSNE_All_Exp} shows the result of computing the MST projected onto the first two principal components computed using our entire experimental dataset of $N_{\text{scans}} = 139$ image plate scans taken from $N_{\text{exp}} = 67$ different experiments. The point size indicates the particular imager modality, while color indicating signal-to-noise ratio (SNR). The data included in this work span a variety of modalities including single- and dual-crystal imager configurations (see Tbl.~\ref{tab:geomparams} for list of imagers with acronyms and single- dual- or orthogonal-imaging configurations fielded). In general, these allow for the selection of particular spectral features to assess physics such as liner-wall or laser entrance hole mix.~\cite{Harding_Prep}. With just two components, we see no indication of clustering based on the imager modality. In this work, we will not consider detailed aspects of different imager modalities that may impact image metrics based on physics of the experiment. Instead, we will focus only on the effects of discrepancies in resolution across the different modalities on image metrics (see Sec.~\ref{sec:res_sens}). However, we do see that there appears to be a region on the right, highlighted by the black dashed ellipse in Fig.~\ref{fig:tSNE_All_Exp} ,which contains the majority of the images having low SNR. This is the first aspect we will explore in Sec.~\ref{sec:Results}, and will point to choices in statistical noise subtraction.

\begin{table*}
%\begin{tabular}
\renewcommand{\arraystretch}{2.5}
\begin{center}
\begin{tabular}{|c|c|c|}
\hline
Imager  & Configuration & Resolution  [$\mu$m$^2$]  \\ \hline
Argon Imager(Ar-Imager) & single &  $15 \times 85$  \\ \hline
Continuum X-ray Imager (CXI) & single &  $59 \times 83$  \\ \hline
High Resolution Continuum X-ray (CXI) & single & $15 \times 16$  \\ \hline
Dual Continuum X-ray (DCXI) & dual  & Ch1 $54 \times 120$  \phantom{...}Ch2 $46 \times 84$ \\ \hline
Iron K-$\alpha _1$ (IKA1) & dual & Ch1 $79 \times 82$  \phantom{...}Ch2 $64\times 66$ \\ \hline
Iron Helium-$\beta$ (IHEB) & dual & Ch1 $63 \times 66$  \phantom{...}Ch2 $50\times 53$ \\ \hline 
Cobalt He-w (CHEWI) & dual/orthogonal & Ch1 $61 \times 66$  \phantom{...}Ch2 $73\times 72$\\ \hline
\end{tabular}
\caption{\label{tab:geomparams}Table giving the full name for imager acronyms as well as specifying whether the imager is single- or dual-crystal in nature. We also provide estimates of the imager resolution. See Ref.~\onlinecite{Harding_Prep} for details. }
\end{center}
%\end{ruledtabular}

\end{table*}

\subsection{\label{sec:noise_sens} Statistical Noise Sensitivity}

In Ref.~\onlinecite{Lewis_JPP_2022}, it was observed that upon subtracting the slowly varying background signal, the non-strand region of SCXI data appeared to obey gaussian noise statistics. We could therefore in-principle utilize a normally distributed noise model to augment our dataset to asses the impact of SNR on the MST. However, here we wish to avoid, to the greatest extent possible, reference to any models when performing data augmentation to assess sensitivities. This will ensure that choices about metric design are as strongly grounded in observation as is feasible. Fortunately, for many of our experiments, multiple scans of the same image plate are available. In the scanning process, the data recorded by exposing the image plates in our experiments is digitized. Briefly, this is achieved by placing the image plate in a scanner which rasterizes a laser over the plate and records the resulting emission from de-excitation of atoms placed into a meta-stable configuration by exposure from the collected x-rays. This process inherently destroys the information contained on the plate. It is possible to saturate the scanner so that in some cases multiple scans of a particular image plate must be obtained to ensure an accurate signal is obtained. We will demonstrate that, when quantifying our images, this aspect is important to account for, at least for the chosen MST metric that we apply, as the destructive scanning process will reduce the signal-to-noise ratio upon sequential scanning. We also note that all of the images utilized in this manuscript used a scanner with pixel resolution of $15~\mu$m $\times 15~\mu$m. Images are cropped to a fixed size of $1.51$~cm wide by $6.825$~cm tall in image coordinates, which corresponds to about $2.5$~mm wide by $1.2$~cm tall in object space coordinates.

\begin{figure}
\includegraphics[width=0.5\textwidth]{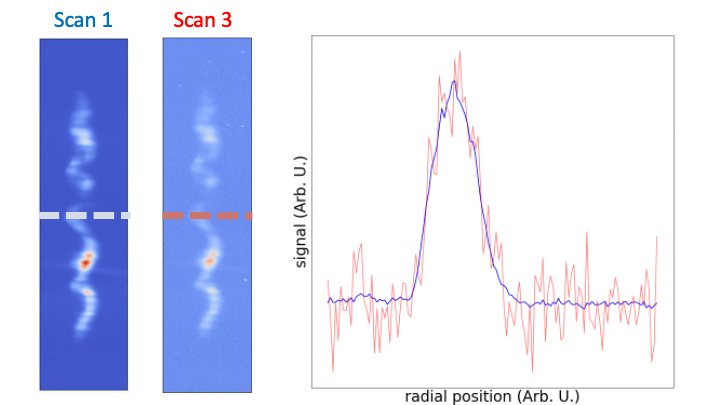}
\caption{Two scans of the same image plate show effectively identical morphology and contrast (peak signal above the floor) as seen by comparing the scans and line outs in the right hand plot. The primary difference is the present of a statistical texture or noise in the red curve from Scan 3 leading to a lower SNR as compared to Scan 1.}
\label{fig:scan_compare}
\end{figure}

Fig.~\ref{fig:scan_compare} demonstrates the impact of multiple scans on a single image plate. The images are nominally identical in important structure provided that the noise doesn't swamp important signal. In general, each sequential scan monotonically decreases SNR, while the initial value is determined by a variety experimental outcomes. It is well known that the MST and other multi-scale wavelet-based metrics are highly sensitive to statistical texture.~\cite{Bruna_IEEE_2013, Allys_PRD_2020} Indeed, in our case, this results in significant contamination of the metric due solely discrepancy in SNR caused by the noise statistics. In what follows, we present two possible approaches to removing this sensitivity. 

\begin{figure}
\includegraphics[width=0.45\textwidth]{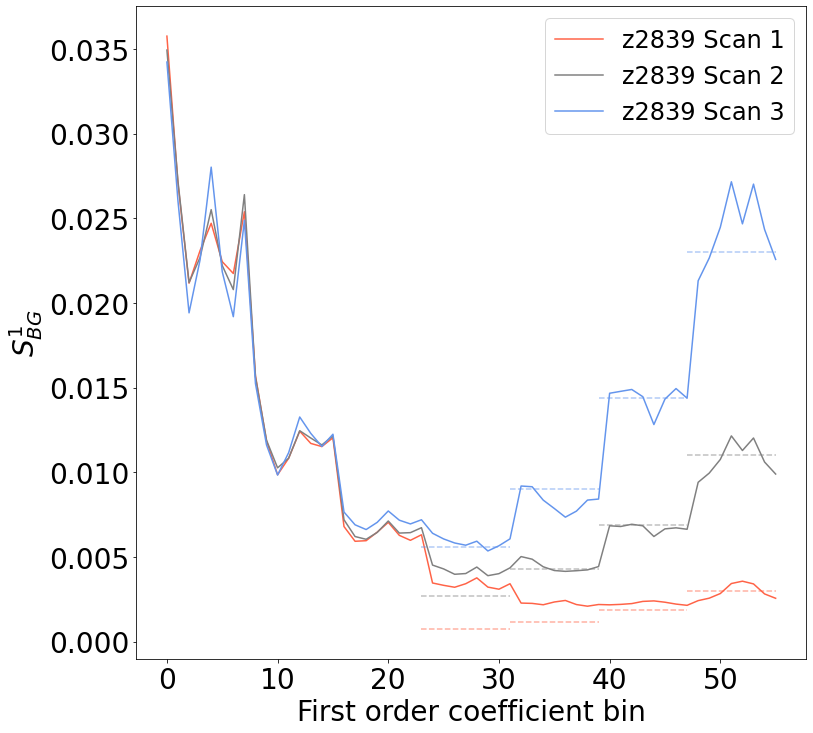}
\caption{The off-strand signal contains contributions from the central strand that have been allowed to ``leak" into the off-strand signal by the father-wavelet as well as a gaussian white-noise like spectrum (horizontal dashed lines). The contribution of noise is monotonically increasing with scan number, corresponding to a monotonically decreasing SNR.}
\label{fig:BG_MST}
\end{figure}

If we consider the first-order off-strand MST coefficients as defined by
\begin{eqnarray}
	S_{\text{BG}} \equiv \dfrac{S_{\text{left}} + S_\text{right}}{2},
\end{eqnarray}  
where $S_{\text{left}(\text{right})}$ are the MST coefficients taken from the left (right) side of the strand, we can gain some insight into the impact of noise statistics on the MST. Fig.~\ref{fig:BG_MST} shows these coefficients for three different scans of the same image plate from $z2839$, where we have normalized to the three low frequency bins according to $\sum_{j = 4}^6 S_{j, \text{BG}}^1 = 1$. Two main features are immediately apparent. First, the low-frequency peak on the left side of the plot corresponds to signal which ``leaks" from the strand into the off strand bins. Second, for the higher scan numbers, there is a very obvious rise again towards high-frequencies. Given that gaussian noise has a uniform power spectrum and our band-pass filter bandwidths scale as $2^{-j}$, one would expect that, for gaussian noise, smaller $j$, or higher frequency would contain a greater contribution from the noise term. We include the expected scaling of gaussian noise for the particular set of band-pass filters used in computing the MST (bandwidths scale specifically as $1.25 \times 2^{-j}$) as dashed lines in Fig.~\ref{fig:BG_MST} where we estimate the overall height for the dashed line by eye using the highest frequency $j-$bin and the subsequent dashed line heights are computed according to the expected scaling. Note that deviation from the scaling only occurs when the amplitude associated with the noise spectrum approaches the level of the ``leaky" strand contribution. The second-order coefficients $S_{\text{BG}}^2$ show similar if slightly more complex features, so we don't show them here. This provides additional evidence along with Ref.~\onlinecite{Lewis_JPP_2022} that gaussian noise statistics is appropriate in describing our data. 

If we assume that the strand information is predominantly low-frequency, then taking into account the fact that the the noise spectrum will be predominantly high-frequency suggests that we may remove the statistical texture by computing
\begin{eqnarray}
	\Delta S = \overline{S-S_\text{BG}} \phantom{.}.\label{eq:offstrandsub}
\end{eqnarray} 
Notice that neither $S$ nor $S_\text{BG}$ are normalized before subtraction and that we perform the spectral normalization given by Eq.~\ref{eq:specnorm} only after performing the subtraction to ensure the noise spectrum does not contribute to the normalization factor. We refer to the procedure prescribed by Eq.~\ref{eq:offstrandsub} as ``off-strand texture subtraction". 

\begin{figure*}
\includegraphics[width=\textwidth]{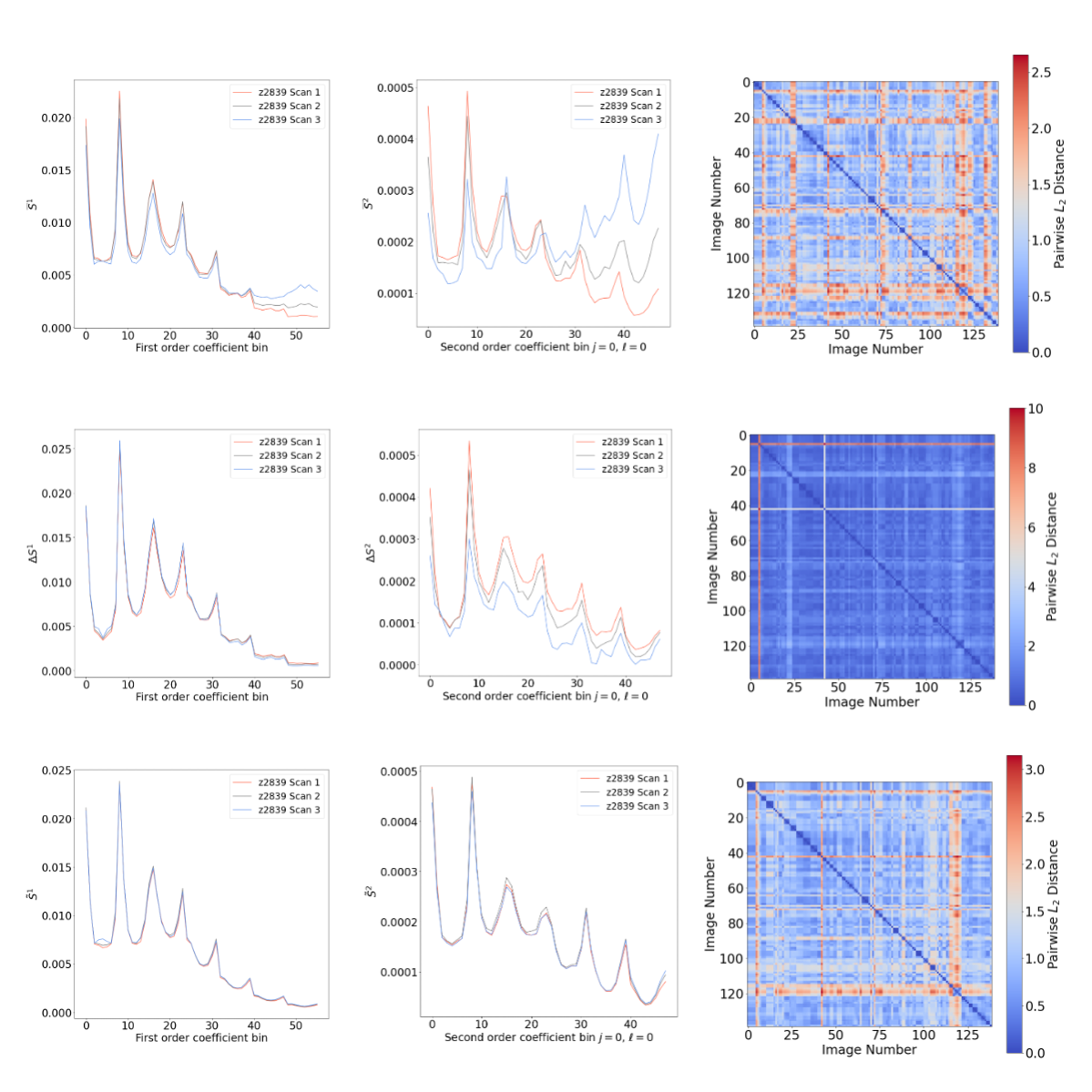}
\caption{Each row has as its first column the first-order MST, second column is the second-order MST, and third column is the corresponding distance matrix. Top row: Original data without texture subtraction showing that the MST differs based on SNR. Middle row: off-strand texture subtraction in the MST space demonstrating good performance on the selected image, but the distance matrix shows that several outliers were generated by the procedure, as discussed in the text. Bottom row: real-space denoising texture subtraction, showing excellent agreement among scans, and only minor discrepancies in the spectral structure.}
\label{fig:texture_subtraction}
\end{figure*}

The first and second rows of Fig.~\ref{fig:texture_subtraction} depict the MST coefficients taken from the center of the strand for $z2839$ (\textit{i.e.} the central pink dot in Fig.~\ref{fig:padding}) for three sequential scans of the same image plate without and with off-strand texture subtraction in the first two columns, while the distance matrix for all experimental data is shown in the right most column. It is apparent from the figure that the process successfully removes the impact of noise in the first-order MST coefficients, but leaves significant variance in the second-order coefficients. We see from the image that at least two significant outliers are generated by the off-strand subtraction procedure.

The most obvious outlier in the off-strand subtracted dataset demonstrated both weak signal from the stagnating fuel and significant background signal that could not be successfully removed using the procedure in Ref.~\onlinecite{Lewis_JPP_2022}. As a result, there were off-strand MST coefficients of significantly larger amplitude than is typical. For this reason, we explore the possibility of denoising our images before computing the MST coefficients.

\begin{figure*}[ht!]
\includegraphics[width=\textwidth]{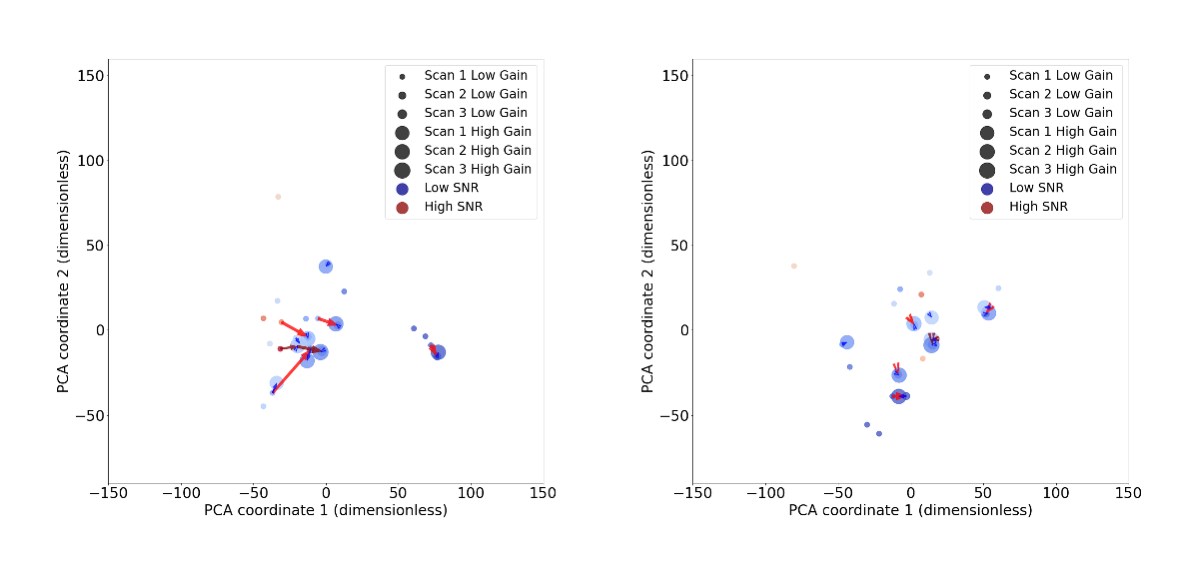}
\caption{Both plots are restricted to CXI data which has the greatest number of multiple scans of image plates. Right: Without applying our ``texture subtraction" procedure, multiple scans of the same image plate can span a range in the metric space that is of a similar size or larger than experiment-to-experiment variations. This is easy to see when considering the red arrows connecting different scans of the same image plate, with the series of dark red arrows that connect the different scans of $z2839$ shown in Fig.~\ref{fig:scan_compare}. Left: By denoising images before computing the MST, we can largely remove the impact of the statistical noise, emphasizing experiment-to-experiment variations that we wish to be primarily due to morphological differences.}
\label{fig:subtraction_PCA}
\end{figure*}

We choose to denoise our $512~\times~512$ images by performing median filtering.~\cite{Gonzales_2017} We note that a similar approach has been used on gated x-ray detector measurements made on implosions in cryogenic Hohlraums at the National Ignition Facility.~\cite{Kyrala_RSI_2010} This will of course result in a loss of image resolution as it blurs the images. The filtering is accomplished by computing a new image from taking the median pixel value over pixels in a disk radius of 7 pixels. However, due to the properties of the MST discussed in Sec.~\ref{sec:MST}, we don't expect the effect to cause a significant change to the strand signal which is dominated by longer wavelengths. The third row of Fig.~\ref{fig:texture_subtraction} shows the result of applying this process, indicating that it is more robust to the features which caused problems for the off-strand subtraction. In addition, it appears that the coefficients are not strongly modified from the first scan, indicating that the approach is effective at removing the impact of noise without drastically altering important features in the strand. The resulting (spectrally normalized) coefficients are denoted by $\tilde S$. 

As previously mentioned, outliers can dominate the dataset variance, severely biasing the resulting principal component vectors. While it is possible to prune the outliers generated by the off-strand texture subtraction, the real-space denoising texture subtraction approach does not suffer from this issue. We therefore  select it as our method for constructing an image metric based on the MST. Note that moving forward any discussion of unsubtracted or subtracted datasets refers to whether or not we work with the original or real-space denoised data respectively. 

Returning to a global PCA perspective, Fig.~\ref{fig:subtraction_PCA} shows the first two principal components restricted now to the CXI modality which has the greatest number of multiple scans of image plates among the different modalities. The left plot shows the data without texture subtraction, while the right is the texture subtracted version, where PCA is computed after performing the texture subtraction. The observed collapse of the red arrows connecting multiple scans of the same image, with the darker red arrows corresponding to the particular case of $z2839$ shown in Fig.~\ref{fig:scan_compare}, indicates that we have removed to a large extend the sensitivity to SNR. We also note that when scanning an image plate, both a low- and high-gain digital image are produced. Provided a high gain doesn't cause the resulting digitized image to be saturated, this corresponds to a simple multiplicative gain. A multiplicative factor will divide out upon normalization of the data, and hence should not show any significant variation. In both the unsubtracted and subtracted cases, there are small blue arrows connecting low and high gain versions of the same scan. There is no observable separation indicating we are not sensitive scaling the images by a multiplicative factor as expected. Finally, we note that although the structure of the distance matrices in Fig.~ \ref{fig:texture_subtraction} for the unsubtracted and denoised cases looks fairly similar, the transformation does significantly alter pairwise distances in a non-trivial way. It is not uncommon for the change to be up to $\pm50\%$ of the original pairwise distance value, and is not uniform across all pairwise relations. As a result, nearest-neighbor relationships can be changed, particularly increasing the likelihood that e.g. $k$-sequential scans of the same image plate will constitute the first $k$-nearest-neighbors to the original scan. This allows us to freely choose the first scan that is not saturated to compare against without concern for SNR causing a large impact when making experiment-to-experiment comparisons.

\subsection{\label{sec:reg_sens} Registration Sensitivity}

\begin{figure*}[ht!]
\includegraphics[width=0.95\textwidth]{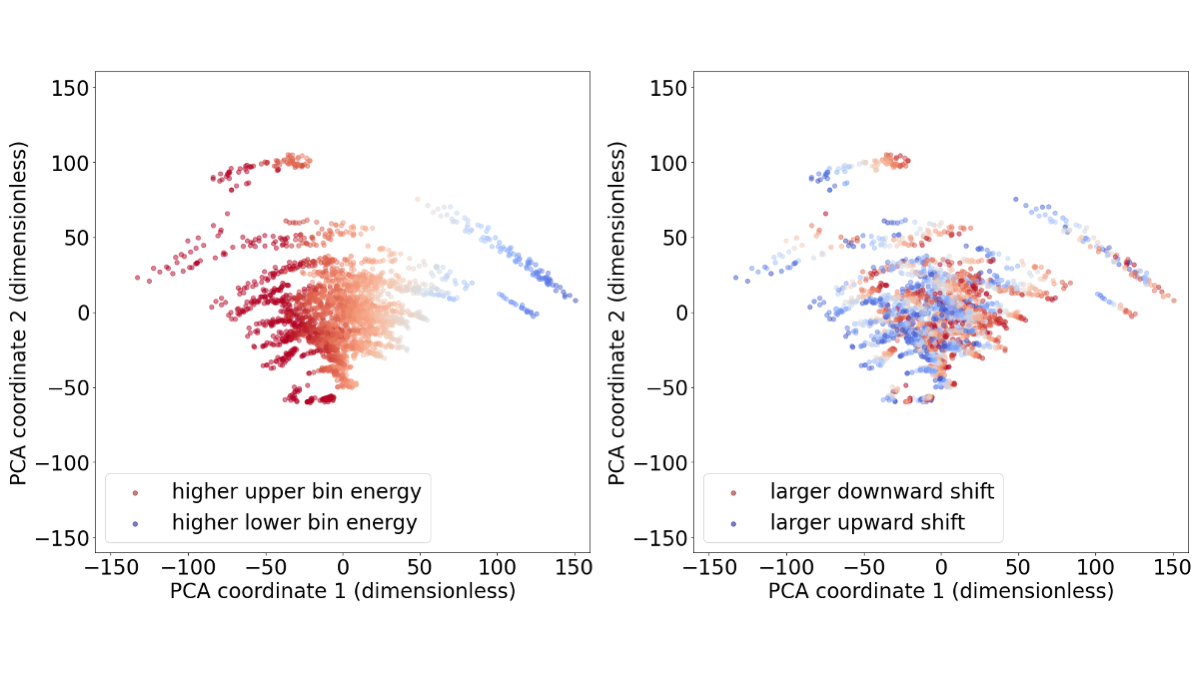}
\caption{We see that the first principal component is largely dependent on the strand asymmetry factor $\Gamma_S$, which in turn depends on the vertical registration of the original image. We should therefore be careful to register images in a reproducible and meaningful fashion.}
\label{fig:transform_sensitivity}
\end{figure*}

Upon further investigation of our dataset, it was discovered that there was a case with residual discrepancy between scans of an otherwise identical image plate after denoising. It turned out that in this case, there was significant plate damage, and slight discrepancies in the original user-specified image crop propagated into differences in the final image registration. Although the MST varies slowly with translation and rotation of the image, it is not invariant to these transformations. Unfortunately, absolute registration would require a spatial fiducial to be fielded on all experiments, is not available for most of the data considered here, and is not an ideal solution in general due to difficulty in practical considerations of fielding a fiducial. In order to assess the impact of translation and rotation, we perform the following data augmentation procedure. First, after denoising, we compute the principal components of the matrix of $\tilde S$ values for our dataset. Second, we randomly rotate between $[-10^\circ,10^\circ]$ about the image center and then apply vertical and horizontal shifts (by integer pixel amounts) up to $\pm20$ pixels, with all transformations performed on the $512 \times 512$ downsampled and denoised images. We then compute $\tilde S$ on this augmented dataset and project onto the principal components computed on the original unaugmented experimental dataset. We show the first two principal components in Fig.~\ref{fig:transform_sensitivity}, with the points in the left image being colored by a vertical asymmetry factor computed as
\begin{eqnarray}
\Gamma_S = \dfrac{\sum \tilde S^1_\text{top} - \tilde S^1_\text{bottom}}{\sum \tilde S^1_\text{middle}},
\end{eqnarray}
while in the right image, the points are colored by the amount of vertical shift applied to the augmented image. We see that there is a strong correlation between $\Gamma_S$ and the first principal component, and a clear trend that shifting images up also pushes the first principal component to more positive values. 

In general, $\Gamma_S$ contains contributions from the inherent strand asymmetry, overall axial length, and any vertical offset that may be present in the image. Scatter plots of the change in first principal component value versus the shift size for each translation and angular value for each rotation appear to be nearly constant with for a given original image. Stated another way, the MST appears to approximately linearize small translations and rotations. While it is not the case here, we note that if the slope of this linear relation were identical across the entire dataset, this would allow projecting the data into the orthogonal compliment of a direction vector associated with the transformation as
\begin{eqnarray}
	v^j_{i}&=& \dfrac{\Delta \text{PC}_i}{\Delta \tau_j},\\
	\hat{\mathbf{v}}^j &=& \dfrac{\mathbf{v}^j}{||\mathbf{v}^j||_2}.
\end{eqnarray}
where $\Delta \text{PC}_i$ is the change in $i^{\text{th}}$ principal component, $\Delta \tau_j$ is the size of a particular transformation applied to the original image. Denoting the orthogonal compliment of $\mathbf{v}$ by $\mathcal{R}_{\mathbf{v},\perp}$, we can project the principal components for each image into this space by computing
\begin{eqnarray}
	\tilde{\mathbf{X}}^n = \mathbf{X}^n - \hat{\mathbf{v}}^j\cdot\mathbf{X}^n \hat{\mathbf{v}}^j \label{eq:projection},
\end{eqnarray}
where $\mathbf{X}^n$ indicates the $d-$dimensional vector representing the principal component decomposition of MST coefficient vector for the $n^\text{th}$ image and $\tilde{\mathbf{X}}^n \in \mathcal{R}_{\mathbf{v},\perp}$. Alternately, one may attempt to choose a summary value that is insensitive to specified transformation to characterize the data. As an example of such a feature, $\alpha \equiv \Delta \Gamma_S/ \Delta y$ is very nearly constant for a given image. The computed value of $\alpha$ characterizes how rapidly the strand energy shifts between the vertical positions sampled along the strand height.

For our purposes, we note that the random transformations do not strongly alter the pairwise neighbor relations to a significant degree. Despite the apparent overlap between experiments seen in Fig.~\ref{fig:transform_sensitivity}, adding even just the third principal component shows that the data form well-separated ``filaments" in the principal component space. Thus for coarse morphological comparison with the MST, small changes to image registration appear to be unimportant. However, in general, detailed comparisons of hand-engineered image features should apply careful consideration of registration sensitivity to understand uncertainties using an analysis similar to the approach shown here.

\begin{figure}[t!]
\includegraphics[width=0.475\textwidth]{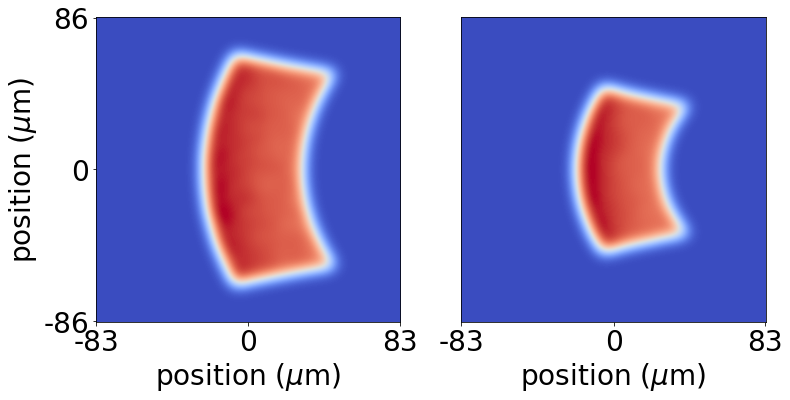}
\caption{Point spread functions for the left- and right-channels of the DCXI configuration. Differential filtering can be applied to enable assessment of \textit{e.g.} attenuation due to liner areal density. It is important to ensure that differences in image metrics due to resolution are not confounded with differences due to spectral filtering between channels.} 
\label{fig:DCXI_PSFs}
\end{figure}

\subsection{\label{sec:res_sens} Resolution Sensitivity}

We claimed earlier that small changes to magnification of the optical system or imager/post-processing induced resolution differences should not strongly impact the MST coefficients. However, the different SCXI imager modalities and even different channels of DCI modalities can exhibit significantly different PSFs. By considering DCXI data, we can uncover the typical scale at which this has impact on the MST. In particular, we note that the optics used to produce HRCXI data have a high-resolution of roughly $15~\mu$m~$\times~15~\mu$m PSF which is on the order of the pixel size for the scanner used to digitize image plates. Note that in practice the PSF and throughput do vary somewhat across the spherical-crystal field-of-view,~\cite{Harding_Prep} an effect we consider to be relatively unimportant for the current demonstration. Therefore, we should be able to effectively treat HRCXI data as a type of ``full-resolution" ground-truth to which different PSFs that further reduce the image resolution can be applied. In Fig.~\ref{fig:DCXI_PSFs} the PSFs obtained from ray-tracing are shown for the left- and right-channel of the DCXI configuration, indicating a fairly significant difference in resolution, especially along the vertical dimension.

\begin{figure}[t!]
\includegraphics[width=0.475\textwidth]{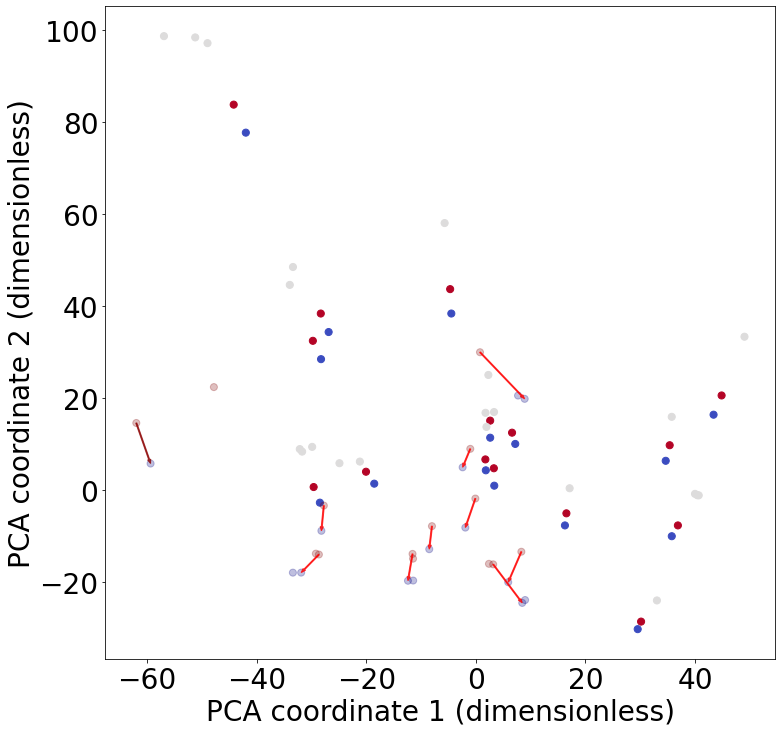}
\caption{By imitating the DCXI dataset by applying left- and right-channel PSFs to the HRCXI dataset, we see that there appears to be a systematic trend associated with the resolution difference. Unfortunately, the trend isn't sufficiently regular and linear over the range of the data, and cannot be projected out using the method proposed in Sec.~\ref{sec:reg_sens}.} 
\label{fig:DCXI_imitation}
\end{figure}

\begin{figure}[t!]
\includegraphics[width=0.475\textwidth]{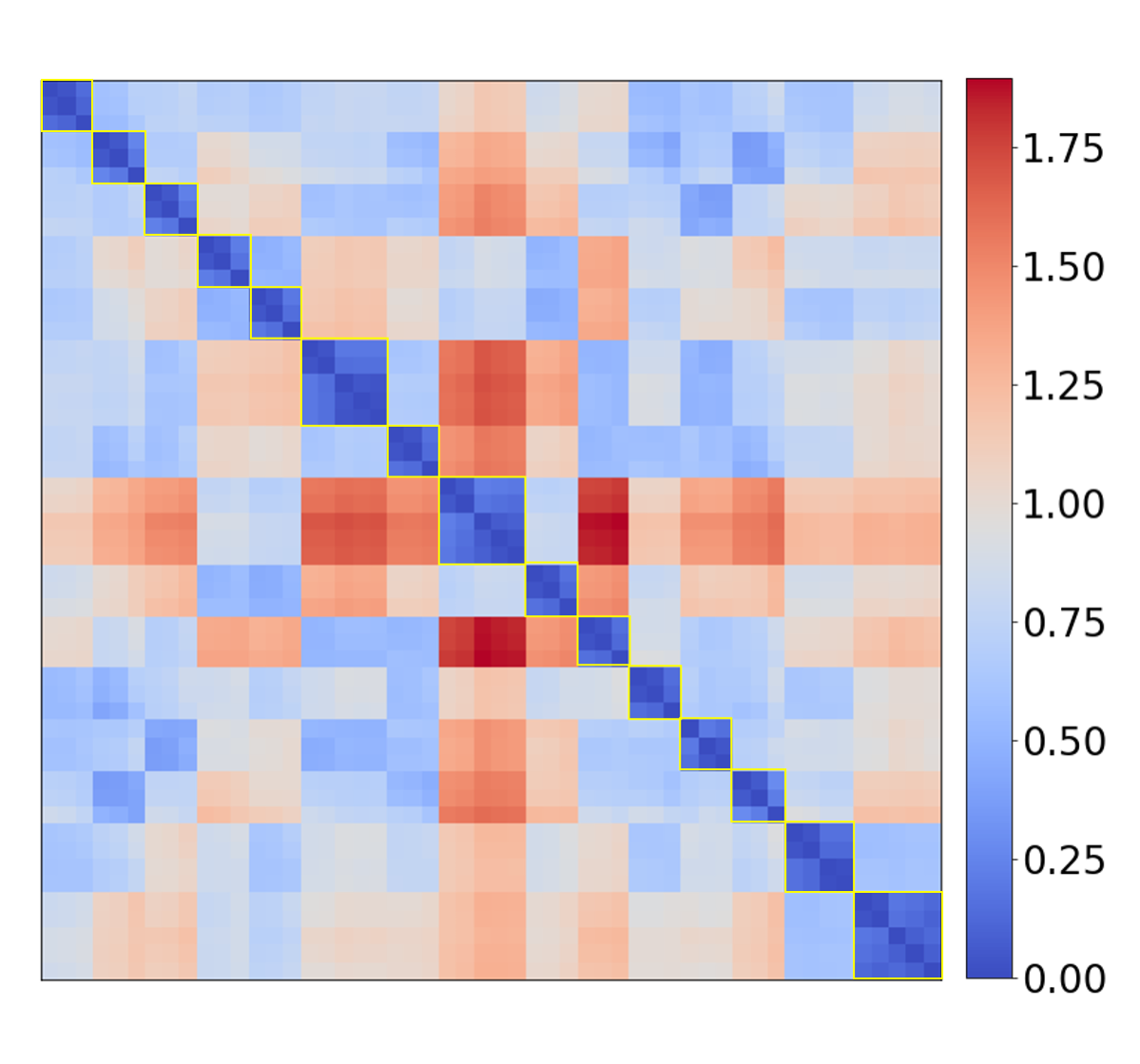}
\caption{The distance matrix for the original HRCXI along with augmentation using the DCXI left- and right-channel point spread functions. The dark blue blocks on the diagonal each contain the two channels in the upper-left along with between one and three corresponding HRCXI images (multiple scans/gain settings) in the remainder of the block. Notice that the left- and right-channels are both significantly closer to the corresponding HRCXI image than to other experimental data. This seems to indicate that the MST is relatively insensitive, though not invariant to changes due to resolution difference between imager modality.} 
\label{fig:DCXI_imitation_dist_mat}
\end{figure}

Fig.~\ref{fig:DCXI_imitation} shows HRCXI data as gray points, with the red and blue points corresponding to convolution with the right- and left-channel DCXI PSFs shown in Fig.~\ref{fig:DCXI_PSFs} respectively. Also shown are the original DCXI data with a red arrow connecting the right- to the left-channel data. We see that both the augmented data and the experimental data show a directional trend pointing generally downward. Unfortunately, neither the augmented nor original data show a constant shift in principal components when going from the right- to left-channel data. As a result, we cannot project out this resolution difference with confidence. However, consideration of the distribution of the difference between left- and right-channel data in principal component space shows that the augmented and experimental data appear to follow a very similar distribution. For the case of DCXI data, one experiment $z3041$ was fielded with differential filtering, left- and right-channel images are connected with the dark red arrow shown in Fig.~\ref{fig:DCXI_PSFs}. The data for $z3041$ does not appear to show significant difference between the two channels in comparison to the distribution of separations in the left- and right-channel images. We also highlight that the two-component view of the data risks biasing our interpretation. For example, it is not obvious from Fig.~\ref{fig:DCXI_imitation} that pairs of left- and right-channel images points tend to be significantly closer together than the between-experiment distance as Fig.~\ref{fig:DCXI_imitation_dist_mat} demonstrates. Fig.~\ref{fig:DCXI_imitation_dist_mat} shows that in general the difference between left- and right-channel images differing only by resolution is quite small, and that on-average, the corresponding HRCXI image bears more resemblance to the lower-resolution augmented versions than with other experimental images. It therefore appears that resolution difference are likely less important than between-experiment morphology differences. However, where possible, it is advisable to ensure that when making detailed inferences, for example regarding apparent wavelengths of morphological structures, that images of identical instrument response are being compared. Alternately, applying the corresponding PSF to HRCXI data should allow direct comparison to continuum-emission data from other imager modalities.

\section{\label{sec:morphology} Morphological Similarity: A Case Study}

\begin{figure*}[ht!]
\includegraphics[width=0.95\textwidth]{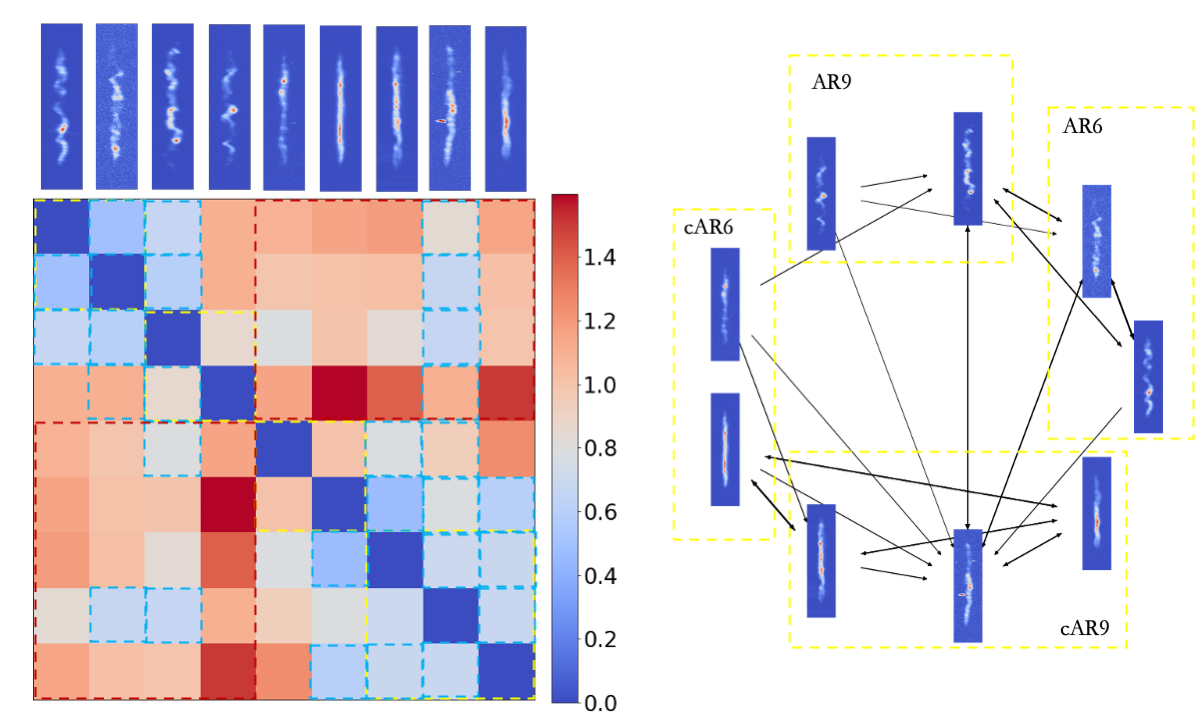}
\caption{Case study demonstrating that the MST is effective at separating images with apparent large scale morphological differences. Left: The pair-wise distance matrix between images using the MST metric. From left to right, the shot numbers are z2839, z2979, z3018, z3303, z2965, z2966, z3019, z3075, and z3135. Right: A graph constructed by allowing a directed edge going from a source image to target image if the target is in the first three nearest neighbors of the source, with the edge weight corresponding to the inverse of the pairwise distance. The nearest neighbors of a source image are indicated by blue dashed boxes in the distance matrix, where the row corresponds to source image and column to target image. Inspection of the structure of the distance matrix and corresponding nearest-neighbor indicates that the two-visually distinct classes are well separated in the MST metric space.} 
\label{fig:Morph_Case_Study_A}
\end{figure*}

A thorough interpretation of the MST in terms of detailed image features goes well beyond the scope and aim of this manuscript. Such an endeavor would involve either reference to a specific geometric model for the fuel plasma such as those presented in Refs.~\onlinecite{Glinsky_PoP_2020,Knapp_Submitted,Lewis_JPP_2022}, and/or significant hand-labeling of image features for our dataset. However, the data-augmentation based approach to assessing and, where possible, removing instrument- and fielding-based effects seems to indicate that the MST provides a useful description of SCXI images. In particular, it appears that the MST is reasonably sensitive to at least gross morphological features. In this section we present an experimental case study in support of this claim. 

The left hand side of Fig.~\ref{fig:Morph_Case_Study_A} shows the distance matrix for a subset of shots that clearly qualitatively fall into two visually separate classes, one where the stagnation is more uniform both in terms of axial intensity variation as well as in terms of the helical amplitude of the strand, and one where it is less uniform in these features. We will label these classes $c$ and $u$ respectively. The dashed red boxes on the distance matrix highlight the pairwise distances between members of these visually distinct classes, while within a row, the blue dashed boxes highlight the first three nearest neighbors to the particular image associated with that row. The yellow dashed boxes highlight groupings of images that have the same liner aspect ratio $AR = r_o/(r_o - r_i)$ , where $r_o$ is the outer radius of the liner and $r_i$ is the inner radius and, if present, dielectric coating (labeled as $cAR$) applied to the outer liner surface. In general, there are fewer blue boxes in the upper-right and lower-left off-diagonal regions indicating that nearest neighbors tend to be within the same ``class". We can visually represent this using a directed graph where an arrow points from one image to another of the target image is in the first three nearest neighbors to the original image as shown in the right column of Fig.~\ref{fig:Morph_Case_Study_A}. The arrow weights in the figure correspond to the inverse of the distance between images in the MST metric space. Inspection of the graph shows that with high probability, transitions between images will tend to keep one within the same class $u$ or $c$. 

\begin{figure*}[t!]
\includegraphics[width=0.95\textwidth]{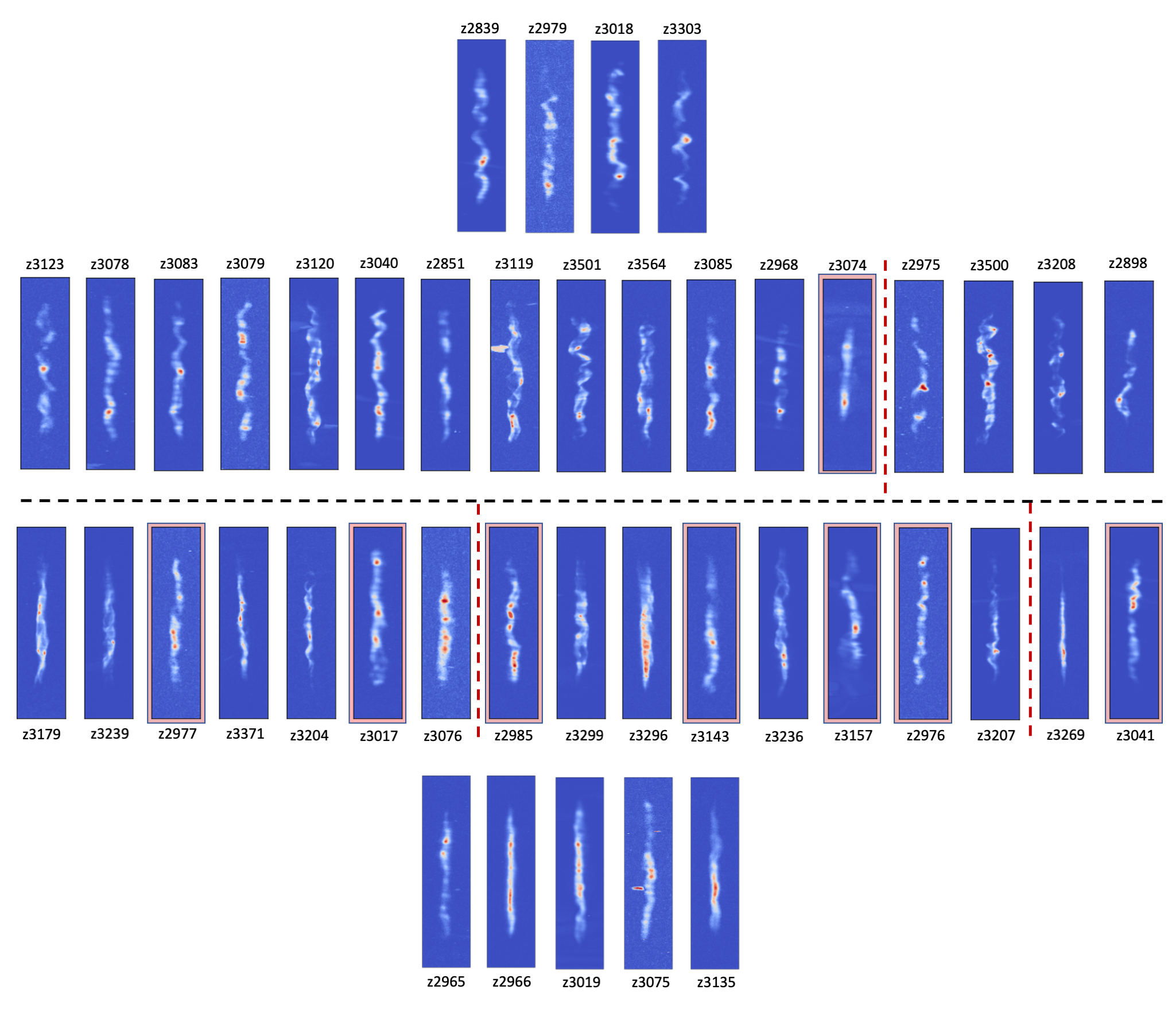}
\caption{First row: Base cases of uncoated targets used to identify if new image is labeled $\tilde u$. Second row: Images labeled as $\tilde u$ as described in the text. Distance from the representative $u$ examples in Fig.~\ref{fig:Morph_Case_Study_A} increases from left to right. Images to the left of the red vertical dashed line have average distance $<0.75$. Red boarded images are misclassified. Third row: Images labeled as $\tilde c$ as described in the text. Distance from the representative $c$ examples in Fig.~\ref{fig:Morph_Case_Study_A} increases from left to right. Images to the left of the first red vertical dashed line have average distance $<0.75$ and the second $<1.0$. Recall that a distance of $1$ corresponds to the median across all images. Fourth row: Base cases of coated targets used to identify if new image is labeled $\tilde c$} 
\label{fig:groups}
\end{figure*}

There are a number of reasonable approaches to characterizing the connectivity of the graph, which by nature of the how the graph was constructed, also characterizes the relationship between the classes $u$ and $c$. We choose to characterize the structure by computing the following probabilities: given that we randomly select an image in class $\alpha$, what is the probability that (assuming transitions among nearest neighbors are equally probable), what is the probability of ending in class $\beta$. It is not difficult to see from Fig.~\ref{fig:Morph_Case_Study_A} that the probabilities are given by
\begin{eqnarray}
P(u \to u) & = & 0.\overline6\overline6,\\
P(u \to c) & = & 0.\overline3\overline3,\\
P(c \to u) & = & 0.2\\
P(c \to c) & = & 0.8.
\end{eqnarray}
Clearly the separation of images spanning between classes is significantly greater than the separation of images within a class, indicating that the MST is able to effectively separate images based on distinct large-scale morphological features. For a detailed discussion of the cause of the morphological difference between stagnation images of experiments using coated and uncoated liners, we refer the reader to Ref.~\onlinecite{Ampleford_Prep}, which provides a detailed quantitative analysis. We also note that the above approach to indicate separation of the classes $u$ and $c$ is qualitatively similar to the classification of images using the MST undertaken in Ref.~\onlinecite{Glinsky_PoP_2020}. However, here we work solely with experimental data rather than a geometric model for the stagnated fuel plasma. Furthermore, the approach using a graph with edges gathered from nearest-neighbor relationships was not pursued in Ref.~\onlinecite{Glinsky_PoP_2020}. The graph-based approach will tend to de-emphasize the absolute magnitude of distance between points. While this is not important when using a parametrized model capable of generating an arbitrarily large dataset, considering the curse of dimensionality in relation to the the relatively high-dimensional MST space along with the relatively small amount of experimental data available, we anticipate that applying the graph-based approach to experimental data may reveal complimentary structure that is not readily apparent based on the distance matrix. 

Going a step further, we may treat these two groupings as representative and label other images based on their average distance to images in each of these groups. Fig.~\ref{fig:groups} shows the first $17$ images labeled for each group, with the upper row being group $u$ and the lower row being group $c$. A total of 17 images were labeled $c$ while 41 were labeled $u$. Taking shots that were dielectric coated to belong to group $c$ while uncoated shots were taken to belong to group $u$ as our ground truth labeling we can obtain type I and type II error and true positive and negative rates. Of the 58 images labeled by this process, a total of 15 were obtained from experiments using coated targets. Thus with probability $P(c) \approx 0.26$ a given image will be an experiment with a coated target. We can then compare the result to that expected from random labeling with the correct \textit{apriori} probability giving
\begin{eqnarray}
&\begin{tabular}{c|c|c} 
random & $u$  & $c$   \\ \hline
$\tilde u$ & $0.74 \pm 0.06$   & $0.74 \pm 0.11$ \\ \hline
$\tilde c$ & $0.26 \pm 0.06$  & $0.26 \pm 0.11$ \\ \hline
\end{tabular} \\
&\begin{tabular}{c|c|c} 
MST & $u$  & $c$   \\ \hline
$\tilde u$ & 0.84 & 0.47  \\ \hline
$\tilde c$ & 0.16 & 0.53 \\ \hline
\end{tabular} 
\end{eqnarray}
where the tilde indicates the label assigned based on proximity to the group and non-tilde quantities indicate the true label and uncertainty in the random case was obtained by taking the standard deviation of $N=10^5$ Monte Carlo samples with a truth label set having 15 positive cases. The use of the MST metric to measure proximity to representative examples does much better than random ($\gtrsim 2\sigma$), and upon considering Fig.~\ref{fig:groups}, would also likely significantly outperform a human labeler. We also considered the possibility of obvious confounding factors. For example, all but two of the coated target images were collected on the HRCXI. Although the results of Sec.~\ref{sec:res_sens} provides strong evidence to the contrary, could resolution explain the labeling performance? Of our 58 images being labeled, 23 were HRCXI images. As a result, we may compute $P(c | \textit{HRCXI}) \approx  0.57$. If one were to label only HRCXI images randomly with the appropriate probability, and label all other images as $\tilde u$, a similar performance is achieved. If our approach reproduced this behavior, it would indicate the possibility of grouping based on resolution, which again we have provided evidence to the counter for in Sec.~\ref{sec:res_sens}. Furthermore, none of the training images in Fig.~\ref{fig:Morph_Case_Study_A} used the HRCXI. In addition, none 7 of the images labeled incorrectly as belonging to the group $\tilde c$ were from the HRCXI indicating that non-HRCXI images are eligible to be labeled as $\tilde c$. Taken together, this provides strong evidence that resolution did not play a significant role in determining the behavior of the classifier. We also emphasize that the classification task marginalizes over a number of important features for individual experiments such as liner aspect ratio, preheat energy deposited, applied magnetic field, current delivery, or fill density of the fuel, which may have additional impact on the morphology. While our dataset size is too small to allow for conditioning on these quantities, the good classification performance without using this information indicates that the applied coating does appear to impact morphology in a similar fashion across experiments with significant discrepancies in input conditions. We refer the interested reader to Ref.~\onlinecite{Ampleford_Prep} for further details on the expected impact of dielectric coatings on morphology, which appears consistent with the findings of this manuscript. 

\section{Conclusion and Future Work}

We have demonstrated the use of a data-augmentation procedure to assess sensitivities of image metrics to realistic features of experimental datasets without reference to a model. In particular, for the MST metric considered here, we demonstrated that the statistical texture created by noise should be carefully treated. On the other hand, small shifts and rotations of an image caused by imperfect image registration as well as difference in resolution across different SCXI imaging modalities does not appear to drastically alter pairwise relationships between points using this metric. Using the MST, we were able to show that two qualitatively distinct classes of image morphology were effectively separated in the metric space. Interestingly, treating these two visually distinct classes as representative examples of the morphology produced in experiments of dielectric coated and uncoated targets allowed for a significantly better than random classification of whether a given experiment used a coated target based purely on morphology, a task which is apparently very difficult by eye. This may hint at additional features that are difficult to quantify. We expect that as more data become available and our ability to simulate dielectric coated targets improves, that these insights can be further explored. 

Our approach to experimental-data-driven assessment of image metrics provides a framework for ensuring that image analysis approaches are not highly sensitive to realistic experimental features. In particular, when constructing hand-selected image features, the work here provides a straightforward way to visualize and understand such sensitivities. In a future publication, we will explore use of the MST and other image metrics to understand the extent to which image structure correlates to physical quantities of interest. 

\begin{acknowledgments}
Sandia National Laboratories is a multimission laboratory managed and operated by National Technology and Engineering Solutions of Sandia LLC (NTESS), a wholly owned subsidiary of Honeywell International Inc., for the U.S. Department of Energy's National Nuclear Security Administration (NNSA) under Contract No. DE-NA0003525. This paper describes objective technical results and analysis. Any subjective views or opinions that might be expressed in the paper do not necessarily represent the views of the U.S. Department of Energy or the United States Government. W. Lewis would like to thank Greg Dunham for useful discussions regarding image digitization. SAND2023-01239O
\end{acknowledgments}

\section*{Data Availability Statement}

The data that support the findings of this study are available from the corresponding author upon reasonable request.

\section{References}
%\bibliography{MagLIF_Image_Metric}% Produces the bibliography via BibTeX.
\bibliographystyle{ieeetr}

\end{document}